\documentclass[letterpaper,preprintnumbers,prd,twocolumn,nofootinbib,nobibnotes,showpacs]{revtex4}
\usepackage{amsfonts}
\usepackage{mathrsfs}
\usepackage{epsfig}
\usepackage{graphicx}%
\usepackage{dcolumn}
\usepackage{amsmath}
\usepackage{color}
\usepackage{color}
\makeatletter
\def\btt#1{\texttt{\@backslashchar#1}}%
\DeclareRobustCommand\bblash{\btt{\@backslashchar}}%
\makeatother
\begin{document}

\title{Exact black hole solutions with non-linear electrodynamic field }
\author{Shuang Yu}\email{yushuang@nao.cas.cn} \affiliation{ Key Laboratory of Computational Astrophysics, National Astronomical Observatories, Chinese Academy of Sciences, Beijing 100012, China}
\affiliation{School of Astronomy and Space Sciences, University of Chinese Academy of Sciences,
No. 19A, Yuquan Road, Beijing 100049, China}\author{Changjun Gao}\email{gaocj@bao.ac.cn}\affiliation{ Key Laboratory of Computational Astrophysics, National Astronomical Observatories, Chinese
Academy of Sciences, Beijing 100012, China}
\affiliation{School of Astronomy and Space Sciences, University of Chinese Academy of Sciences,
No. 19A, Yuquan Road, Beijing 100049, China}

\date{\today}

\begin{abstract}
We construct exact black hole solutions to Einstein gravity with nonlinear electrodynamic field. In these solutions, there are in general four parameters. They are physical mass, electric charge, cosmological constant and the coupling constant. Due to the presence of coupling constant, these solutions differ significantly from the Reissner-Nordstrom-de Sitter solution in Einstein-Maxwell gravity with a cosmological constant. In the first place, some of them are endowed with a topological defect on angle $\theta$. Secondly, for the extreme charged black holes, the electric charge can be much larger or much smaller than the mass by varying the coupling constant. Thirdly, for very large coupling constant, the radius of event horizon for some black holes approaches $2M$ regardless of the electric charge. Lastly, some black holes are asymptotically neither flat nor de Sitter (or anti-de Sitter).
\end{abstract}



\maketitle

\section{Introduction}
\label{sec:1}

The nonlinear electrodynamics is firstly proposed by Born and Infield in 1934 in order that the
self-energy of a point-like charge \cite{BI1:1934,BI2:1934} is finite . Afterwards, there were not many studies on nonlinear electromagnetic
fields. Until the 1980s, it is found that the effective action for the open string ending on D-branes can exactly be written in the nonlinear form \cite{brane:1,brane:2,brane:3}. These huge discoveries make people aggrandize the devotion to study nonlinear electrodynamics in the aspect of cosmology by a long way \cite{nl:0,nl:1,nl:2,nl:3,nl:4,nl:5,nl:6,nl:7,nl:8,nl:9,nl:10,nl:11,nl:12,nl:13,nl:14,nl:15,nl:16,nl:17,nl:18,nl:19,nl:20,nl:21,nl:22,nl:23}.
It is then found that if the early universe is dominated by the nonlinear electromagnetic field, the initial Big-Bang singularity can be avoided.
Not only that, a period of inflation of the universe can subsequently be achieved.
Furthermore, with the help of AdS/CFT correspondence \cite{ads:00}, the nonlinear electrodynamics string approach has been
applied to obtain solutions describing baryon configurations. Even more interesting than these findings, are that nonlinear electromagnetic fields can play the role of cosmic dark energy.

On the other hand, from the point of view of black hole theory, it is interesting to seek for
the exact solutions of Einstein gravity plus nonlinear
electromagnetic field. The reason is that these solutions may help to understand the
physical relevance of nonlinear effects in strong gravitational and strong electromagnetic fields.
In this aspect, some solutions of charged black holes and black branes \cite{solu:1,solu:2,solu:3,solu:4,solu:5,solu:6,solu:7,solu:8,solu:9,solu:10,solu:11,solu:12,solu:13,solu:14,solu:15,solu:16,solu:17,solu:18,solu:19,solu:20}, magnetic black holes and magnetic branes \cite{solu:21,solu:22,solu:23} have been found. Except for in general relativity, the solutions in higher derivative gravity with nonlinear electromagnetic fields have also been studied in the literature \cite{solu:24,solu:25,solu:26,solu:27,solu:28}.
It is found that not only the Big-Bang singularity but also the black hole singularity can be avoided by using the non-linear electromagnetic field. Correspondingly, some regular black hole solutions without singularities are obtained \cite{solu:29,solu:30,solu:31,solu:32,solu:33,solu:34,solu:35,solu:36,solu:37,solu:38,solu:39,solu:40}. As for the black hole horizons, the Kerr-Newman-de Sitter spacetime has at most three horizons, the inner horizon, the event horizon and the cosmic horizon. By taking account of non-linear electromagnetic field, one can construct black hole spacetime with as many horizons as desired \cite{mul:1,mul:2,mul:3,mul:4,mul:5}. However, the method adopted in some researches, for example \cite{mul:1,mul:2,mul:3,mul:4,mul:5} is questionable.

According to the conventional method, one always start from the Lagrangian of the Einstein gravity together with the nonlinear electromagnetic field. Given the Lagrangian, then the equations of motion are derived. By solving the equations of motion, we obtain the solution for the metric field and the electric field. The integration constants in the solution are conserved quantities and they are understood as the physical mass, electric charge and so on. However, in the preceding researches, one assumes the metric initially and then solves for the electromagnetic field and the Lagrangian function. This is an upside-down method. Following this method, the constants assumed in the metric would inevitably present in the Lagrangian. This makes the Lagrangian very ugly. More seriously, beginning from the derived
Lagrangian, one would inevitably obtain different solutions from the initial assumed ones because of the presence of integration constants.

Thus the purpose of this paper is to construct exact black hole solutions for Einstein gravity with nonlinear electrodynamic field by using the conventional method. In order to find the exact solutions, we have examined several Lagrangians. In this paper we shall present six exact solutions. In these solutions, there are in general four parameters. They are physical mass, electric charge, cosmological constant and the coupling constant. The physical mass and electric charge take their place as integration constants. Due to the presence of coupling constant, these solutions differ significantly from the Reissner-Nordstrom-de Sitter solution in Einstein-Maxwell gravity with a cosmological constant. For example, some black holes are endowed with a topological defect on angle; the electric charge of extreme black holes can not be  equal to the mass; the black hole spacetime can be asymptotically neither flat nor de Sitter (or anti-de Sitter).

The paper is organized as follows. In Sec. II, we derive
the equations of motion in the theories of nonlinear
electrodynamics. In Sec. III, we present six  black hole
spacetimes and give an analysis on their causal structure. Then Sec. IV gives the conclusion and discussion. Throughout this paper, we adopt the system of units in which $G=c=\hbar=1$ and the metric signature
$(-,\ +,\ +,\ +)$.
\section{equations of motion}
\label{sec:2}
The action of nonlinear electrodynamic theories which are minimally coupled to gravity is
\begin{equation}
S=\frac{1}{16\pi}\int\sqrt{-g}\left[R+K\left(\psi\right)\right]d^4x\;,
\end{equation}
with
\begin{equation}
\psi=F_{\mu\nu}F^{\mu\nu}\;,\ \ \ \  F_{\mu\nu}=\nabla_{\mu}A_{\nu}-\nabla_{\nu}A_{\mu}\;.
\end{equation}
Here $R$ is the Ricci scalar and $A_{\mu}$ is the Maxwell field. $K(\psi)$ is the function
of $\psi$. The variation of the action with respect to the metric gives
the Einstein equations
\begin{eqnarray}
G_{\mu\nu}=-2K_{,\psi}F_{\mu\lambda}F_{\nu}^{\lambda}+\frac{1}{2}g_{\mu\nu}K\;, \ \ \ \ \ K_{,\psi}\equiv\frac{dK}{d\psi}\;.
\end{eqnarray}
The variation of the action with respect to the field $A_{\mu}$ gives the generalized
Maxwell equations
\begin{eqnarray}
\nabla_{\mu}\left(K_{,\psi}F^{\mu\nu}\right)=0\;.
\end{eqnarray}
In the background of static and spherically symmetric spacetime which can
always be parameterized as
\begin{eqnarray}
ds^2=-U\left(r\right)dt^2+\frac{1}{U\left(r\right)}dr^2+f\left(r\right)^2d\Omega_{2}^{2}\;.
\end{eqnarray}
Here $d\Omega_2^2= d\theta^2+\sin^2\theta d\phi^2$. Due to the spacetime is static and spherically symmetric, the non-vanishing component of Maxwell field $A_{\mu}$ is uniquely to be
\begin{eqnarray}
A_0=\phi\left(r\right)\;,
\end{eqnarray}
and $\psi$ is
\begin{eqnarray}
\psi=-2\phi^{'2}\;,
\end{eqnarray}
by resorting to a gauge transformation of $A_{\mu}\rightarrow
A_{\mu}+\nabla_{\mu}\chi$.
Then we obtain the Einstein equations and the generalized Maxwell equation
\begin{eqnarray}
-\frac{U^{'}f^{'}}{f}-\frac{2Uf^{''}}{f}+\frac{1}{f^2}-\frac{Uf^{'2}}{f^2}&=&2K_{,\psi}\phi^{'2}+\frac{1}{2}K\;, \label{7}\\
-\frac{U^{'}f^{'}}{f}+\frac{1}{f^2}-\frac{Uf^{'2}}{f^2}&=&2K_{,\psi}\phi^{'2}+\frac{1}{2}K\;, \label{8}\\
\frac{U^{'}f^{'}}{f}+\frac{Uf^{''}}{f}+\frac{1}{2}U^{''}&=&-\frac{1}{2}K\;, \label{9}\\
\left(f^2K_{,\psi}\phi^{'}\right)^{'}&=&0\;.\label{10}
\end{eqnarray}
The prime denotes the derivative with respect to $r$. Eqs.~(8-10) come from $G_0^0=\rho$, $G_1^1=p_r$ and $G_2^2=p_{\theta}$, respectively. Eq.~(11) is
the equation of motion for the Maxwell field and it gives the definition of the electric charge,
\begin{eqnarray}
Q \equiv f^2K_{,\psi}\phi^{'}\;.
\end{eqnarray}

Due to the Bianchi identities, only three of the four equations are independent. So we have only three unknown functions among $U$, $f$, $\phi$ and $K$.
As usual, we assume the expressions of $K(\psi)$ initially. Then we are left with three unknown
functions, $U$, $f$ and $\phi$ and the system of equations are closed.

In the next, we present $6$ exact black hole solutions by investigating $6$ different expressions of $K(\psi)$.
Before the presentation, we observe the difference of Eq. (8) and Eq. (9) which gives
\begin{eqnarray}
f^{''}=0\;.
\end{eqnarray}
Thus we obtain the physical solution for $f$,
\begin{eqnarray}
f=r\;.
\label{12}
\end{eqnarray}
Therefore, the metric of a static and spherically symmetric spacetime with the source of nonlinear electrodynamic field is
\begin{eqnarray}
ds^2=-U\left(r\right)dt^2+\frac{1}{U\left(r\right)}dr^2+r^2d\Omega_{2}^{2}\;.\label{5}
\end{eqnarray}
The horizons of the spacetime are located at
\begin{eqnarray}\label{horizons}
U=0\;.
\end{eqnarray}

\section{Black hole solutions}\label{sec:3}
In this section, we derive $6$ exact black hole solutions.
\subsubsection{$K=-2 \sqrt{2} \alpha \sqrt{-\psi}+\psi+2 \Lambda$}
In this Lagrangian, $\alpha$ is a coupling constant and it has the dimension of inverse of some length, ${L_{a}}$. In order that the $\alpha$ term can be safely neglected in the scale $L_{e}$ which is experimentally interested, we should have  $L_e\ll L_a$.  Since $\psi$ is negative because of Eq.~(7), there is a minus under the square root. When $\alpha=0$, it reduces to the Maxwell term $\psi$ plus the cosmological constant term $2\Lambda$.

We find the solution is given by
\begin{eqnarray}
\phi&=&-\alpha r-\frac{Q}{r}\;,\\
U&=&1-\frac{2 M}{r}+\frac{Q^{2}}{r^{2}}-\frac{1}{3} r^{2} \Lambda-\frac{1}{3} r^{2} \alpha^{2}+2 Q \alpha\;.
\end{eqnarray}
Here $M$ and $Q$ are the black hole mass and charge, respectively. In the case of vanishing $\alpha$, it is nothing but the Reissner-Nordstrom-de Sitter solution. If $\alpha\neq 0$, two terms $-\frac{1}{3} r^{2} \alpha^{2}$ and $2 Q \alpha$ arise in the metric. But since $L_e\ll L_a$, the two terms can be safely neglected.  The $r^{2} \alpha^{2}$ term  makes the spacetime asymptotically de Sitter provided that $\Lambda=0$ and $2 Q \alpha\ll 1$. On the other hand, the presence of term $2 Q \alpha$ endows the spacetime a topological defect. In order to understand this point, let $\Lambda=-\alpha^2$ and $r\rightarrow \infty$, then the spacetime is asymptotically
\begin{eqnarray}
ds^2&=&-\left(1+2Q\alpha\right)dt^2+\frac{1}{1+2 Q \alpha}dr^2+r^2d\Omega^2\;.
\end{eqnarray}
Rescale $t,r,\theta,\phi$, we find the metric can be written as
\begin{eqnarray}
ds^2&=&-dt^2+dr^2+r^2\left[d\theta^2+sin^2\left(\omega\theta\right)d\phi^2\right]\;,
\end{eqnarray}
with
\begin{eqnarray}
\omega&=&\frac{1}{\sqrt{1+2\alpha Q}}\;.
\end{eqnarray}
Thus the three dimensional space is endowed with a topological defect on angle $\theta$. We note that, in general, one always have $2\alpha Q\ll 1$.

Same as the Reissner-Nordstrom-de Sitter solution, the spacetime has at most three horizons. One is the de Sitter horizon and the others are the outer event horizon and the inner Cauchy horizon, respectively.

\subsubsection{$K=\psi-\frac{1}{3}\alpha\sqrt2{(-\psi)^{(3/2)}}+2\Lambda$}
In this Lagrangian, $\alpha$ has the dimension of length, ${L_{a}}$. In order that the $\alpha$ term can be safely neglected in the scale $L_{e}$ which is experimentally interested, we should have  $L_e\gg L_a$. This is different from the first solution where $L_e\ll L_a$. When $\alpha=0$, it reduces to the Maxwell theory $\psi$ plus the cosmological constant term $2\Lambda$.

\begin{figure}[h]
\begin{center}
\includegraphics[width=9cm]{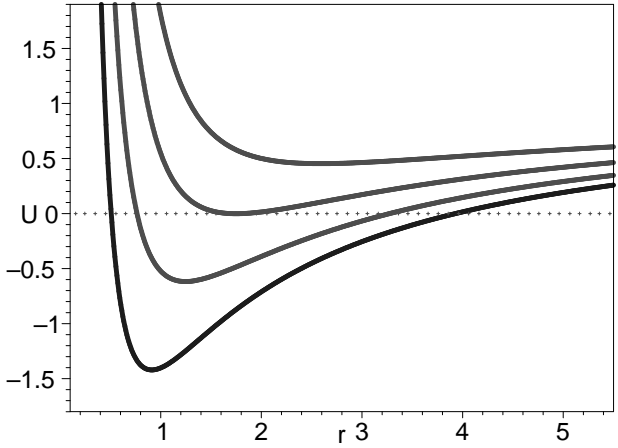}
\caption{The $U-r$ relation with different electric charges
$Q=1.92,1.76,1.6,1.44$ from up to down. It shows that with the increasing of electric charge $Q$, the outer event horizon shrinks while the inner Cauchy horizon expands. When $Q=1.76$, is an extreme black hole. If $Q>1.76$, we have a naked singularity. We have put $M=1$ and $\alpha=0.03$.
}\label{fig:urq2}
\end{center}
\end{figure}

We find the solution is given by
\begin{eqnarray}\label{u2}
U&=&1-\frac{2 M}{r}-\frac{1}{3} \Lambda r^{2} +\frac{r^{2}}{18\alpha^{2}}+\frac{Q}{\alpha}\nonumber\\&&-\frac{1}{18} \frac{\left(r^{2}+4 \alpha Q\right)^{\frac{3}{2}}}{r \alpha^{2}}-\frac{2}{3} \frac{Q \sqrt{r^{2}+4 \alpha Q}}{r \alpha}\nonumber\\&& +\frac{4 Q^{2} }{3 r \sqrt{\alpha Q}}\ln \left(\frac{8 \alpha Q+4 \sqrt{\alpha Q} \sqrt{r^{2}+4 \alpha Q}}{r}\right)\;,
\end{eqnarray}
\begin{eqnarray}
\phi&=&\frac{Q}{\sqrt{\alpha Q}}\ln \left(\frac{8 \alpha Q+4 \sqrt{\alpha Q} \sqrt{r^{2}+4 \alpha Q}}{r}\right)\nonumber\\&&
 +\frac{\sqrt{r^{2}+4 \alpha Q}}{2\alpha}- \frac{r}{2\alpha}\;.\label{Ap}
\end{eqnarray}
The expression of $U$ tells us $\alpha$ should have the same sign with the electric charge in order that the square root is always meaningful.
Expanding the solution in the series of $r$, we obtain
\begin{eqnarray}
U&=&1-\frac{2M}{r}+\frac{Q^2}{r^2}-\frac{1}{3}\Lambda r^2+\frac{2}{3}\frac{Q^2\ln\left(16\alpha Q\right)}{\sqrt{\alpha Q}r}\nonumber\\&&
 -\frac{2\alpha Q^3}{9r^4}+\frac{\alpha^2 Q^4}{5r^6}-\frac{2\alpha^3 Q^5}{7r^8}+\textrm{O}\left(\frac{1}{r^{10}}\right)\;.
\end{eqnarray}
Since $L_e\gg L_a$, the $\alpha$ terms can be dropped compared to other terms. Thus it restores to the Reissner-Nordstrom-de Sitter solution. For small $\alpha$, $U$ can be written as
\begin{eqnarray}
U&=&1-\frac{2M}{r}+\frac{Q^2}{r^2}-\frac{1}{3}\Lambda r^2-\frac{2M^{'}}{r}\;.
\end{eqnarray}
with
\begin{eqnarray}
M^{'}&\equiv&-\frac{1}{3}\frac{Q^2\ln\left(16\alpha Q\right)}{\sqrt{\alpha Q}}\;.
\end{eqnarray}
So the effect of a small coupling constant $\alpha$ amounts to an effective mass. Then we have $Q=M+M^{'}$ for the extreme black hole. This means the electric charge $Q$ can be much larger than the black hole mass $M$ for the extreme black hole in the absence of $\Lambda$. To show this point more clearly, we plot the $U-r$ relation in Fig.~(\ref{fig:urq2}) by using Eq.~(\ref{u2}). In Fig.~(\ref{fig:urq2}), we plot the $U-r$ relation with different electric charges $Q=1.92,1.76,1.6,1.44$ from up to down. It shows that with the increasing of electric charge $Q$, the outer event horizon shrinks while the inner Cauchy horizon expands. When $Q=1.76$, it becomes an extreme black hole. If $Q>1.76$, we have a naked singularity. We let $M=1$, $\Lambda=0$ and $\alpha=0.03$ here.

\begin{figure}[h]
\begin{center}
\includegraphics[width=9cm]{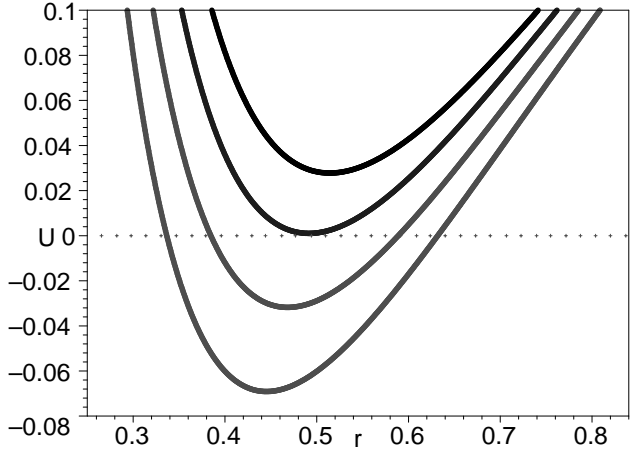}
\caption{The $U-r$ relation with different coupling constants
$\alpha=0.7,0.8,0.9,1.0$ from up to down. It shows that with the decreasing of coupling constant $\alpha$, the outer event horizon shrinks while the inner Cauchy horizon expands. When $\alpha=0.8$, is an extreme black hole. If $\alpha<0.8$, we have a naked singularity. We have put $M=1$ and $Q=0.65$.
}\label{fig:ur2}
\end{center}
\end{figure}

In the case of small $\alpha$, there is in general two horizons in this spacetime (with $\Lambda=0$). One is the outer event horizon and the other is the inner Cauchy horizon. This is the same as the Reissner-Nordstrom spacetime. However, if $\alpha$ is not very small, the situation is interesting. As an example, in Fig.~\ref{fig:ur2}, we plot the $U-r$ relation with different coupling constants $\alpha=0.7,0.8,0.9,1.0$ from up to down. From the figure, we see that with the decreasing of coupling constant $\alpha$, the outer event horizon shrinks while the inner Cauchy horizon expands. When $\alpha=0.8$, it becomes an extreme black hole. If $\alpha<0.8$, we have a naked singularity. We put $M=1$ and $Q=0.65$ in this situation. We note that the black hole can be extreme for $Q=0.65$ which is smaller than the black hole mass $M$.

In all, the electric charge $Q$ of the extreme black hole can be much larger or much smaller than the black hole mass $M$. The exact values of charge $Q$ depends on the coupling constant $\alpha$.

\subsubsection{$K=\psi-\frac{1}{4} \alpha \psi^{2}+2 \Lambda$}

Here $\alpha$ has the dimension of square of length, ${L_{a}}$. If $L_e\gg L_a$, the $\alpha$ term can be neglected. This is different from the first solution where $L_e\ll L_a$.

\begin{figure}[h]
\begin{center}
\includegraphics[width=9cm]{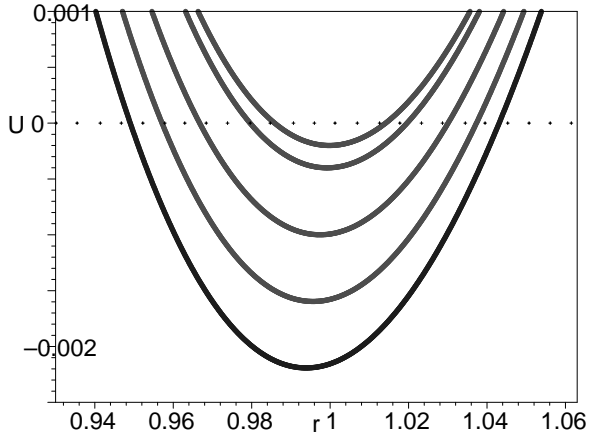}
\caption{The $U-r$ relation with different coupling constants
$\alpha=0,0.002,0.008,0.014,0.02$ from up to down. It shows that with the increasing of coupling constant $\alpha$, the outer event horizon expands while the inner Cauchy horizon shrinks. We have put $M=1$ and $Q=0.9999$.
}\label{fig:ur3}
\end{center}
\end{figure}

We find the solution is given by
\begin{eqnarray}
\phi&=&\frac{12^{\frac{1}{3}}}{6\alpha }\int \frac{\left(\gamma^{\frac{2}{3}}-12^{\frac{1}{3}} \alpha r^{2}\right)}{ r\gamma^{\frac{1}{3}}} dr\;,\\
U&=&1-\frac{2 M}{r}-\frac{1}{3} \Lambda r^{2}-\frac{1}{9} \frac{r^{2}}{\alpha}-\frac{3^{\frac{2}{3}}Q}{2\alpha r} \int \frac{\left(\alpha^{2} r \Gamma\right)^{\frac{1}{3}}}{\gamma^{\frac{1}{3}} r}dr\nonumber\\&&+\frac{ 18^{\frac{1}{3}}}{3r}\int \frac{r\left(\Gamma-\alpha r^{5}\right)}{\gamma^{\frac{4}{3}}} d r
\nonumber\\&&+\frac{ 3^{\frac{2}{3}}}{3r} \int \frac{r^{4}\left(\alpha^{2} r \Gamma\right)^{\frac{1}{3}}}{\gamma^{\frac{4}{3}}} d r\;,
\end{eqnarray}
where
\begin{eqnarray}
&&\gamma \equiv \left(9 Q+\sqrt{3} \sqrt{\frac{4 r^{4}+27Q^{2}\alpha}{\alpha}}\right) \alpha^{2} r\;,\nonumber\\&&
\Gamma \equiv 3 Q \gamma+2\alpha r^{5}\;.
\end{eqnarray}
The expression of $\gamma$ tells us $\alpha$ should be positive in order that the square root makes sense. $U$ is rather involved. So we expand it in the series of $r$,
\begin{eqnarray}
U&=&1-\frac{2 M}{r}+\frac{Q^2}{r^2}-\frac{1}{3} \Lambda r^{2}-\frac{\alpha Q^4}{10r^6}\nonumber\\ &&+\frac{\alpha^2Q^6}{9r^{10}}-\frac{3\alpha^3 Q^8}{13r^{14}}+\textrm{O}\left(\frac{1}{r^{18}}\right)\;.
\end{eqnarray}
We see the solution restores to the Reissner-Nordstrom-de Sitter solution when $\alpha=0$. Furthermore, it is asymptotically flat when $\Lambda=0$.
In Fig.~(\ref{fig:ur3}), by fixing $M=1, \Lambda=0$ and $Q=0.9999$,  we plot the $U-r$ relation with different coupling constants
$\alpha=0,0.002,0.008,0.014,0.02$ from up to down. The figure shows, with the increasing of coupling constant $\alpha$, the inner Cauchy horizon shrinks while the outer event horizon expands. Thus the presence of coupling constant is equivalent to the decreasing of electric charge of Reissner-Nordstrom black hole.

\subsubsection{$K=\psi+2 \alpha \ln (-\psi)+2 \Lambda$}
Here $\alpha$ has the dimension of inverse of length squared, ${L_{a}}$. If $L_e\ll L_a$, the $\alpha$ term can be neglected. This is the same as the first solution.

We obtain the solution as follows
\begin{eqnarray}
\phi&=&\int \frac{Q+\sqrt{Q^{2}+4 r^{4} \alpha}}{2r^{2}}dr\;,\\
U(r)&=&1-\frac{2 M}{r}-\frac{1}{3} r^{2} \Lambda+\frac{Q^{2}}{2r^{2}}+\frac{1}{3} r^{2} \alpha\nonumber\\&&
+\frac{1}{3} r^{2} \alpha \ln 2-\frac{Q}{2r}\int\frac{\sqrt{Q^{2}+4 r^{4} \alpha}}{r^{2}} d r\nonumber\\&&
-\frac{2\alpha }{r}\int r^{2}
\ln \frac{|Q+\sqrt{Q^{2}+4 r^{4} \alpha}|}{r^{2}}dr
\;,
\end{eqnarray}
where the plus and minus in $\pm$ correspond to the positive and the negative charge $Q$, respectively.
It is apparent $\alpha$ should be positive in order that the square root is always meaningful. When $\alpha=0$, the solution restores to the Reissner-Nordstrom-de Sitter solution. In order to see the behavior of $U$ clearly when $\alpha\neq 0$, we expand $U$ in the series of $r$,
\begin{eqnarray}
U&=&\left[\frac{1}{3}\alpha-\frac{1}{3}\alpha\left(\ln 2+\ln \alpha\right)-\frac{1}{3}\Lambda\right]r^2+1-2\sqrt{\alpha}Q\nonumber\\&&-\frac{2M}{r}+\frac{Q^2}{2r^2}+\frac{Q^3}{36\sqrt{\alpha}r^4}
-\textrm{O}\left(\frac{1}{r^8}\right)\;.
\end{eqnarray}
For simplicity, let $\Lambda=0$. Then we conclude that when
\begin{eqnarray}
\alpha>\frac{e}{2}\;,
\end{eqnarray}
the spacetime is asymptotically de Sitter. There are generally two horizons in this spacetime. One of them is the black hole event horizon and the other is the de Sitter horizon. On the other hand, when
\begin{eqnarray}
0<\alpha<\frac{e}{2}\;,
\end{eqnarray}
it is asymptotically anti-de Sitter. In this case, there leaves only one black hole event horizon.
Finally, when
\begin{eqnarray}
\alpha={\frac{e}{2}}\;,
\end{eqnarray}
it is asymptotically flat but with a topological defect on angle $\theta$. For this spacetime, there is only one black hole event horizon provided that
\begin{eqnarray}
0\leq |Q|\leq {\frac{1}{\sqrt{2e}}}\;.
\end{eqnarray}
If
\begin{eqnarray}
|Q|>{\frac{1}{\sqrt{2e}}}\;,
\end{eqnarray}
there would be no horizon and the spacelike singularity is naked.
\subsubsection{$K= -\frac{2}{\alpha^2}\mathrm{e}^{\alpha \sqrt{-\psi}}+2 \Lambda$}\label{sec:4}

Here $\alpha$ has the dimension of length. Expanding $K$ in the series of $\psi$, we have
\begin{eqnarray}
K=2\Lambda-\frac{2}{\alpha^2}-\frac{2}{\alpha}\sqrt{-\psi}+\psi+\frac{1}{3}\alpha \psi\sqrt{-\psi}+\textrm{{O}}\left(\psi^{{2}}\right)\;.
\end{eqnarray}
The second term is the closest equivalent to a minus cosmological constant. The fourth term is the Maxwell one. Since $\psi$ scales approximately in general
\begin{eqnarray}
\psi\sim -\frac{1}{r^2}\;,
\end{eqnarray}
we expect the Maxwell term is dominant in some domain of $r$. Thus the Einstein-Maxwell theory is produced.

\begin{figure}[h]
\begin{center}
\includegraphics[width=9cm]{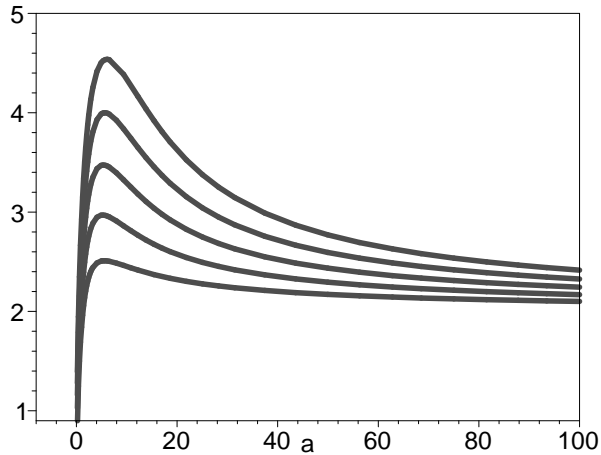}
\caption{The variation of black hole event horizon with respect to the coupling constant $\alpha$ for different charges, $Q=2.4,2.0,1.6,1.2,0.8$ from up to down. It shows that with the increasing of coupling constant $\alpha$, the event horizon first increases and
then it shrinks. In the end, it approaches the Schwarzschild radius $2M$ regardless of the value of charge $Q$. We have put $M=1$.
}\label{fig:ra}
\end{center}
\end{figure}

\begin{figure}[h]
\begin{center}
\includegraphics[width=9cm]{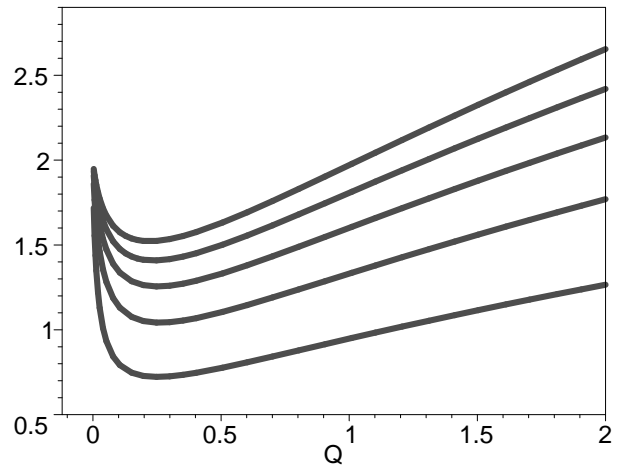}
\caption{The variation of black hole event horizon with respect to the charge $Q$ for different coupling constants, $\alpha=1.0,0.8,0.6,0.4,0.2$ from up to down. It shows that with the increasing of charge $Q$, the event horizon first decreases and
then it increases. On the other hand, with the decreasing of charge $Q$, the event horizon approaches the Schwarzschild radius $2M$ irrespective of the coupling constants. We have put $M=1$.}\label{fig:rq}
\end{center}
\end{figure}

We obtain the solution as follows
\begin{eqnarray}
\phi=\frac{\sqrt{2}r}{4\alpha}\left[\ln 2+2\ln |{\alpha Q}|+4-4\ln r\right]\;,
\end{eqnarray}
\begin{eqnarray}\label{eq:ln}
U(r)&=&\frac{\sqrt{2}Q}{2\alpha}\left[4\ln r-\ln 2-2\ln |{\alpha Q}|-2\right]\nonumber\\&&+1-\frac{2 M}{r}-\frac{1}{3} \Lambda r^{2}\;.
\end{eqnarray}
We are interested in the case of $\Lambda=0$. Without the loss of generality, we consider only the situation of positive charge, $Q>0$. We find when $\alpha< 0$, there are two horizons, namely, the black hole event horizon and the cosmic-like apparent horizon.
When $\alpha\geq 0$, there is only one black hole event horizon. In Fig.~(\ref{fig:ra}), by putting $M=1$, we plot the variation of black hole event horizon with respect to the coupling constant $\alpha$ for different charges, $Q=2.4,2.0,1.6,1.2,0.8$ from up to down. It shows that with the increasing of coupling constant $\alpha$, the event horizon first increases and
then it shrinks. In the end, it approaches the Schwarzschild radius $2M$ regardless of the value of charge $Q$.
Similarly, in Fig.~(\ref{fig:rq}) by putting $M=1$, we plot the variation of black hole event horizon with respect to the charge $Q$ for different coupling constants, $\alpha=1.0,0.8,0.6,0.4,0.2$ from up to down. It shows that with the increasing of charge $Q$, the event horizon first decreases and
then it increases. On the contrary, with the decreasing of charge $Q$, the event horizon approaches the Schwarzschild radius $2M$ irrespective of the coupling constants. This is consistent with Fig.~(\ref{fig:ra}).

\subsubsection{$K=-\frac{1}{\alpha}\mathbf{e}^{-\alpha \psi}+2 \Lambda$}
Here $\alpha$ has the dimension of square of length. Expanding $K$ in series of $\psi$, we have
\begin{eqnarray}
K=-\frac{1}{\alpha}+2\Lambda+\psi-\frac{\alpha}{2}{\psi^2}+\frac{1}{6}\alpha^2 \psi^3+\textrm{{O}}\left(\psi^{{4}}\right)\;.
\end{eqnarray}
The third term $\psi$ is just the Maxwell one.

We obtain the solution as follows
\begin{eqnarray}
\phi&=&\int\frac{1}{2}\sqrt{\frac{\zeta}{\alpha}}dr\;,\\
U&=&1-\frac{2 M}{r}-\frac{1}{3} r^{2} \Lambda\nonumber\\&&
-\frac{Q}{\alpha r}\int{\sqrt{{\alpha}{\zeta}}}dr+\frac{Q}{\alpha r}\int{\sqrt{\frac{\alpha}{\zeta}}}dr\;,\\
\zeta &\equiv& \operatorname{Lambert}\textrm{W}\left(\frac{4 \alpha Q^{2}}{r^{4}}\right)\;.
\end{eqnarray}
Taking into account the definition of Lambert\textrm{W} function and the domain of $r$, we must require $\alpha\geq0$.

\begin{figure}[h]
\begin{center}
\includegraphics[width=9cm]{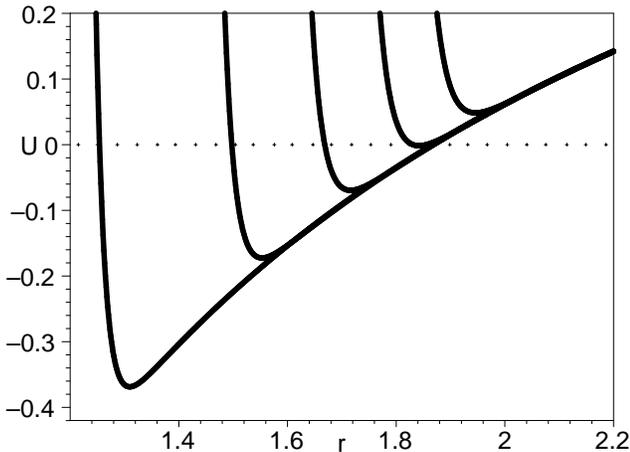}
\caption{The $U-r$ relation with different coupling constants
$\alpha=2,4,6,8,10$ from left to right. It shows that with the increasing of coupling constant $\alpha$, the inner Cauchy horizon expands while the outer event horizon gradually shrinks. When $\alpha=8$, the two horizons coincide and the black hole becomes the extreme one. We have put $M=1$ and $Q=0.5$. }\label{fig:6a}
\end{center}
\end{figure}

\begin{figure}[h]
\begin{center}
\includegraphics[width=9cm]{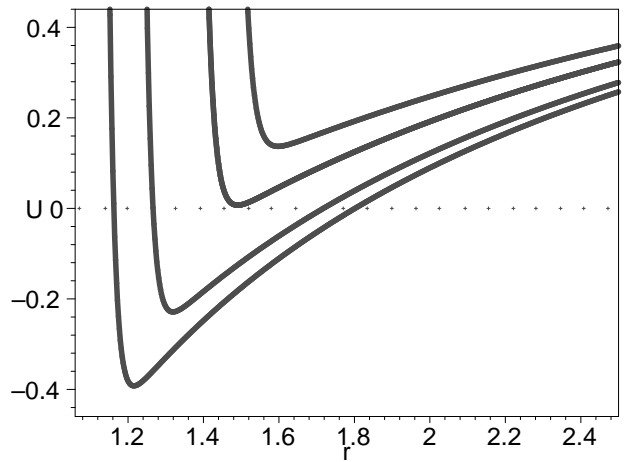}
\caption{The $U$---$r$ relation with different charge
$Q=1,0.88,0.7,0.6$ from up to down. It shows that with the increasing of charge $Q$, the inner Cauchy horizon expands and the outer event horizon shrinks. When $Q=0.88$, the two horizons coincide and the black hole becomes the extreme. When $Q>0.88$, the two horizons disappear and the naked singularity is present.  We have put $M=1$ and $\alpha=1$.}\label{fig:6b}
\end{center}
\end{figure}

In order to understand the relation  between $U$  and $r$ more clearly, we expand $U$ in the series of $r$. Then we obtain
\begin{eqnarray}\label{exp}
U&=&\left(\frac{1}{6\alpha}-\frac{1}{3}\Lambda\right)r^2+1-\frac{2 M}{r}+\frac{Q^2}{r^2}-\frac{\alpha Q^4}{5r^6}\nonumber\\&&+\frac{10\alpha^2Q^6}{27r^{10}}-\frac{49\alpha^3 Q^8}{39 r^{14}}+\textrm{{O}}\left(r^{{-18}}\right) \;.
\end{eqnarray}
It is apparent when
\begin{eqnarray}\label{Lam}
\Lambda&\neq&\frac{1}{2\alpha}\;,
\end{eqnarray}
the spacetime is asymptotically either de Sitter or anti-de Sitter. On the other hand, when
\begin{eqnarray}\label{Lam}
\Lambda&=&\frac{1}{2\alpha}\;,
\end{eqnarray}
the spacetime is asymptotically flat. In this case, we find in general there are two horizons in this spacetime, namely, the outer black hole
event horizon and the inner Cauchy horizon. In Fig.~(\ref{fig:6a}), by fixing $M=1$ and $Q=0.5$, we plot the $U-r$ relation with different coupling constants
$\alpha=2,4,6,8,10$ from left to right. From the figure, we can see that with the increasing of coupling constant $\alpha$, the inner Cauchy horizon expands while the outer event horizon gradually shrinks. When $\alpha=8$, the two horizons coincide and the black hole becomes the extreme one. Subsequently, for $\alpha>8$, the horizons disappear and a naked singularity is present. We note that due to the effect of coupling constant $\alpha$, the black hole can be extreme even if $Q<M$. This is different from the extreme Reissner-Nordstrom black hole where $Q=M$. In Fig.~(\ref{fig:6b}), by fixing $M=1$ and $\alpha=1$, we plot the $U-r$ relation with different charge
$Q=1,0.88,0.7,0.6$ from up to down. The figure shows, with the increasing of charge $Q$, the inner Cauchy horizon expands and  the outer event horizon shrinks. When $Q=0.88$, the two horizons coincide and the black hole becomes the extreme one. When $Q>0.88$, the two horizons disappear and the naked singularity is present.

\section{Conclusion and Discussion}\label{sec:8}
 In conclusion, we have obtained six exact black hole solutions by using the traditional method. Although these solutions can restore to the well-known Reissner-Nordstrom-de Sitter solution when they are expanded in the series of $r$, they have many different properties from the later. In the first place, some of them are endowed with a topological defect on angle $\theta$. In other words, the spacetime is not asymptotically Minkowski spacetime. Secondly, for the extreme charged black holes, the electric charge can be much larger or much smaller than the mass by varying the coupling constants. Thirdly, for very large coupling constant, the radius of event horizon for some black holes approaches $2M$ regardless of the electric charge. This means the effect of electromagnetic field is negligible for large coupling constant. Fourthly, some black holes are asymptotically neither flat nor de Sitter (or anti-de Sitter). Actually, they are asymptotically $\ln r$ (see Eq.~(\ref{eq:ln}) when $\Lambda=0$). Taken into account the weak field limit, the corresponding strength of gravitational field is $\sim\frac{Q}{\alpha r}$. Then the rotation curves of galaxies might be interpreted by using this solution. Finally, the thermodynamics, the gravitational waves as well as other issues of these black holes are not investigated. Bearing in mind these aspects, we plan to carry  out the research elsewhere.

\section*{Acknowledgments}
This work is partially supported by China Program of International ST Cooperation 2016YFE0100300
, the Strategic Priority Research Program ``Multi-wavelength Gravitational Wave Universe'' of the
CAS, Grant No. XDB23040100, the Joint Research Fund in Astronomy (U1631118), and the NSFC
under grants 11473044, 11633004 and the Project of CAS, QYZDJ-SSW-SLH017.

\newcommand\ARNPS[3]{~Ann. Rev. Nucl. Part. Sci.{\bf ~#1}, #2~ (#3)}
\newcommand\AL[3]{~Astron. Lett.{\bf ~#1}, #2~ (#3)}
\newcommand\AP[3]{~Astropart. Phys.{\bf ~#1}, #2~ (#3)}
\newcommand\AJ[3]{~Astron. J.{\bf ~#1}, #2~(#3)}
\newcommand\GC[3]{~Grav. Cosmol.{\bf ~#1}, #2~(#3)}
\newcommand\APJ[3]{~Astrophys. J.{\bf ~#1}, #2~ (#3)}
\newcommand\APJL[3]{~Astrophys. J. Lett. {\bf ~#1}, L#2~(#3)}
\newcommand\APJS[3]{~Astrophys. J. Suppl. Ser.{\bf ~#1}, #2~(#3)}
\newcommand\JHEP[3]{~JHEP.{\bf ~#1}, #2~(#3)}
\newcommand\JMP[3]{~J. Math. Phys. {\bf ~#1}, #2~(#3)}
\newcommand\JCAP[3]{~JCAP {\bf ~#1}, #2~ (#3)}
\newcommand\LRR[3]{~Living Rev. Relativity. {\bf ~#1}, #2~ (#3)}
\newcommand\MNRAS[3]{~Mon. Not. R. Astron. Soc.{\bf ~#1}, #2~(#3)}
\newcommand\MNRASL[3]{~Mon. Not. R. Astron. Soc.{\bf ~#1}, L#2~(#3)}
\newcommand\NPB[3]{~Nucl. Phys. B{\bf ~#1}, #2~(#3)}
\newcommand\CMP[3]{~Comm. Math. Phys.{\bf ~#1}, #2~(#3)}
\newcommand\CQG[3]{~Class. Quantum Grav.{\bf ~#1}, #2~(#3)}
\newcommand\PLB[3]{~Phys. Lett. B{\bf ~#1}, #2~(#3)}
\newcommand\PRL[3]{~Phys. Rev. Lett.{\bf ~#1}, #2~(#3)}
\newcommand\PR[3]{~Phys. Rep.{\bf ~#1}, #2~(#3)}
\newcommand\PRd[3]{~Phys. Rev.{\bf ~#1}, #2~(#3)}
\newcommand\PRD[3]{~Phys. Rev. D{\bf ~#1}, #2~(#3)}
\newcommand\RMP[3]{~Rev. Mod. Phys.{\bf ~#1}, #2~(#3)}
\newcommand\SJNP[3]{~Sov. J. Nucl. Phys.{\bf ~#1}, #2~(#3)}
\newcommand\ZPC[3]{~Z. Phys. C{\bf ~#1}, #2~(#3)}
\newcommand\IJGMP[3]{~Int. J. Geom. Meth. Mod. Phys.{\bf ~#1}, #2~(#3)}
\newcommand\IJMPD[3]{~Int. J. Mod. Phys. D{\bf ~#1}, #2~(#3)}
\newcommand\IJMPA[3]{~Int. J. Mod. Phys. A{\bf ~#1}, #2~(#3)}
\newcommand\GRG[3]{~Gen. Rel. Grav.{\bf ~#1}, #2~(#3)}
\newcommand\EPJC[3]{~Eur. Phys. J. C{\bf ~#1}, #2~(#3)}
\newcommand\PRSLA[3]{~Proc. Roy. Soc. Lond. A {\bf ~#1}, #2~(#3)}
\newcommand\AHEP[3]{~Adv. High Energy Phys.{\bf ~#1}, #2~(#3)}
\newcommand\Pramana[3]{~Pramana.{\bf ~#1}, #2~(#3)}
\newcommand\PTP[3]{~Prog. Theor. Phys{\bf ~#1}, #2~(#3)}
\newcommand\APPS[3]{~Acta Phys. Polon. Supp.{\bf ~#1}, #2~(#3)}
\newcommand\ANP[3]{~Annals Phys.{\bf ~#1}, #2~(#3)}


\begin{thebibliography}{99}
\bibitem{BI1:1934} M. Born and L. Infeld, \PRSLA{143}{410}{1934}.
\bibitem{BI2:1934} M. Born and L. Infeld, \PRSLA{144}{425}{1934}.
\bibitem{brane:1} E. S. Fradkin and A. Tseytlin, \PLB{163}{123}{1985}.
\bibitem{brane:2} A. Tseytlin, \NPB{276}{391}{1986}.
\bibitem{brane:3} N. Seiberg and E. Witten, \JHEP {09}{032}{1999}.
\bibitem{nl:0} M. Novello, S. E. Perez Bergliaffa and J. Salim, \PRD{69}{127301}{2004}.
\bibitem{nl:1} M. Novello and S. E. Perez Bergliaffa, \PR{463}{127}{2008}.
\bibitem{nl:2} M. Novello, E. Goulart, J. M. Salim and S. E. Perez Bergli-
affa, \CQG{24}{3021}{2007}.
\bibitem{nl:3} V. A. De Lorenci, R. Klippert, M. Novello and J. M. Salim,
\PRD{65}{063501}{2002}.
\bibitem{nl:4} M. Novello, Aline N. Araujo and J. M. Salim, \IJMPA{24}{5639}{2009}.
\bibitem{nl:5} C.S. Camara, J. C. Carvalho and M. R. De Garcia Maia,
\IJMPD{16}{427}{2007}.
\bibitem{nl:6} V. V. Dyadichev, D. V. Gal¡¯tsov and P. V. Moniz, AIP
Conf. Proc. 861, 312 (2006)
\bibitem{nl:7} V. V. Dyadichev, D. V. Gal¡¯tsov and P. V. Moniz, \PRD{72}{084021}{2005}.
\bibitem{nl:8} M. Novello, \IJMPA{20}{2421}{2005}.
\bibitem{nl:9} M. Novello, AIP Conf. Proc. 782, 306 (2005)
\bibitem{nl:10} R. Garcia-Salcedo and N. Breton, \CQG{22}{4783}{2005}.
\bibitem{nl:11}C. S. Camara, M. R. de Garcia Maia, J. C. Carvalho and
J. A. S. Lima, \PRD{69}{123504}{2004}.
\bibitem{nl:12} R. Garcia-Salcedo and N. Breton, \CQG{20}{5425}{2003}.
\bibitem{nl:13} E. Elizalde, J. E. Lidsey, S. Nojiri and S. D. Odintsov,
\PLB{574}{1}{2003}.
\bibitem{nl:14} D. N. Vollick, \GRG{35}{1511}{2003}.
\bibitem{nl:15} V. V. Dyadichev, D. V. Gal¡¯tsov, A. G. Zorin and M. Yu.
Zotov, \PRD{65}{084007}{2002}.
\bibitem{nl:16} P. V. Moniz, \PRD{66}{103501}{2002}.
\bibitem{nl:17} P. Moniz, \CQG{19}{L127}{2002}.
\bibitem{nl:18} R. Garcia-Salcedo and N. Breton, \IJMPA{15}{4341}{2000}.
\bibitem{nl:19} B. L. Altshuler, \CQG{7}{189}{1990}.
\bibitem{nl:20}D. N. Vollick, \PRD{78}{063524}{2008}.
\bibitem{nl:21} C. G. Callan, A. Guijosa, K. G. Savvidy and O. Tafjord,
\NPB{555}{183}{1999}.
\bibitem{nl:22} P. Vargas Moniz, \PRD{66}{103501}{2002}.
\bibitem{nl:23} P. Vargas Moniz, \CQG{19}{L127}{2002}.
\bibitem{ads:00}O. Aharony, S. S. Gubser, J. Maldacena, H. Ooguri and
Y. Oz, \PR{323}{183}{2000}.

\bibitem{solu:1}E. Ayon-Beato and A. Garcia, \PRL{80}{5056}{1998}.
\bibitem{solu:2} E. Ayon-Beato and A. Garcia, \PLB{464}{25}{1999}
\bibitem{solu:3}E. Ayon-Beato and A. Garcia, \GRG{31}{629}{1999}.
\bibitem{solu:4} E. Ayon-Beato and A. Garcia, \GRG{37}{635}{2005}.
\bibitem{solu:5} N. Breton, Phys. Rev. D 67, 124004 (2003)
\bibitem{solu:6} R. G. Cai, D. W. Pang and A. Wang, \PRD{70}{124034}{2004}.
\bibitem{solu:7} R. G. Cai and Y. W. Sun, \JHEP{09}{115}{2008}.
\bibitem{solu:8} S. S. Yazadjiev, \PRD{72}{044006}{2005}.
\bibitem{solu:9} D. N. Vollick, \PRD{72}{084026}{2005}.
\bibitem{solu:10} Y. S. Myung, Y. W. Kim and Y. J. Park, \PRD{78}{044020}{2008}.
\bibitem{solu:11} Y. S. Myung, Y. W. Kim and Y. J. Park, \PRD{78}{084002}{2008}.
\bibitem{solu:12} S. H. Mazharimousavi, M. Halilsoy and Z. Amirabi, \PRD{78}{064050}{2008}.
\bibitem{solu:13} A. Khodam-Mohammadi, \GC{15}{154}{2009}.
\bibitem{solu:14}M. Hassaine and C. Martinez, \PRD{75}{027502}{2007}.
\bibitem{solu:15} M. Hassaine and C. Martinez, \CQG{25}{195023}{2008}.
\bibitem{solu:16} H. Maeda, M. Hassaine and C. Martinez, \PRD{79}{044012}{2009}.
\bibitem{solu:17}S. H. Mazharimousavi and M. Halilsoy, \PLB{681}{190}{2009}.
\bibitem{solu:18} M. H. Dehghani and H. R. Rastegar Sedehi, \PRD{74}{124018}{2006}.
\bibitem{solu:19}M. H. Dehghani, S. H. Hendi, A. Sheykhi and H. Rastegar
Sedehi, \JCAP{02}{020}{2007}.
\bibitem{solu:20} S. H. Hendi, \JMP{49}{082501}{2008}.
\bibitem{solu:21} K.A. Bronnikov, \PRD{63}{044005}{2001}.
\bibitem{solu:22} M. H. Dehghani, N. Bostani and S. H. Hendi, \PRD{78}{064031}{2008}.
\bibitem{solu:23} M. H. Dehghani, A. Sheykhi and S. H. Hendi, \PLB{659}{476}{2008}.
\bibitem{solu:24} M. H. Dehghani and S. H. Hendi, \IJMPD{16}{1829}{2007}.
\bibitem{solu:25} S. Mukherji and S. Pal, [arXiv:08062507]
\bibitem{solu:26} M. H. Dehghani and S. H. Hendi, \GRG{41}{1853}{2009}
\bibitem{solu:27} M. Aiello, R. Ferraro and G. Giribet, \PRD{70}{104014}{2004}.
\bibitem{solu:28} M. H. Dehghani, N. Alinejadi and S. H. Hendi, \PRD{77}{104025}{2008}.
\bibitem{solu:29} K. A. Bronnikov, \PRD{63}{044005}{2001}.
\bibitem{solu:30} E. Ayon-Beato and A. Garcia, \PLB{493}{149}{2000}.
\bibitem{solu:31} E. Elizalde and S. R. Hildebrandt, \PRD{65}{124024}{2002}.
\bibitem{solu:32} I. Dymnikova, \CQG{21}{4417}{2004}.
\bibitem{solu:33}P. Nicolini, A. Smailagic and E. Spallucci, \PLB{632}{547}{2006}.
\bibitem{solu:34}S. Ansoldi, P. Nicolini, A. Smailagic and E. Spallucci, \PLB{645}{261}{2007}.
[gr-qc/0612035].
\bibitem{solu:35}S. Hossenfelder, L. Modesto and I. Premont-Schwarz, \PRD{81}{044036}{2010}.
\bibitem{solu:36} T. Johannsen, \PRD{88}{044002}{2013}.
\bibitem{solu:37}I. Dymnikova and E. Galaktionov, \CQG{32}{165015}{2015}.
\bibitem{solu:38} H. Culetu, Acta Phys. Polon. Supp. {\bf 10}, {431}(2017).
\bibitem{solu:39} M. S. Ma, \AP{362}{529}{2015}.
\bibitem{solu:40} G. Kunstatter, H. Maeda and T. Taves, \CQG{33}{105005}{2016}.
\bibitem{solu:41} P. Pradhan, \GRG{48}{19}{2016}.
\bibitem{solu:42} M. E. Rodrigues, J. C. Fabris, E. L. B. Junior and G. T. Marques, \EPJC{76}{250}{2016}.
\bibitem{solu:43}Z. Y. Fan and X.Wang, \PRD{94}{124027}{2016}.
\bibitem{solu:44} S. Chinaglia and S. Zerbini, \GRG{49}{75}{2017}.
\bibitem{mul:1} K. A. Bronnikov, I. G. Dymnikova and E. Galaktionov, \CQG{29}{095025}{2012}.
\bibitem{mul:2}S. V. Bolokhov, K. A. Bronnikov and M. V. Skvortsova, \CQG{29}{245006}{2012}.
\bibitem{mul:3}K. A. Bronnikov, K. A. Baleevskikh and M. V. Skvortsova, \PRD{96}{124039}{2017}.
\bibitem{mul:4}S. Nojiri and S. D. Odintsov, \PRD{96}{104008}{2017}.
\bibitem{mul:5}C. Gao, Y. Lu, S. Yu and Y. Shen, \PRD{97}{104013}{2018}.




\end{thebibliography}
\end{document}